\documentclass[twocolumn]{autart}    % Enable this line and disable the 
\usepackage{graphicx}          % Include this line if your 
\usepackage{amsmath}
\usepackage[english]{babel} 
\usepackage{amssymb}
\usepackage{dsfont}
\usepackage{graphicx}
\usepackage{theorem}
\usepackage{cite}
\usepackage{pstricks}
\usepackage{booktabs}
\usepackage{epsfig}
\usepackage{pst-grad}
\usepackage[normalem]{ulem}
\usepackage{bbm}
\usepackage{float}
\usepackage{footnote}
\usepackage{siunitx}

\DeclareMathOperator{\col}{col}
\DeclareMathOperator{\diag}{diag}

\usepackage[american,smartlabels,siunitx]{circuitikz}
\usepackage[american,siunitx]{circuitikz}

\usetikzlibrary{calc}
\ctikzset{bipoles/thickness=1}
\ctikzset{bipoles/length=0.8cm}
\ctikzset{bipoles/diode/height=.375}
\ctikzset{bipoles/diode/width=.3}
\ctikzset{tripoles/thyristor/height=.8}
\ctikzset{tripoles/thyristor/width=1}
\ctikzset{bipoles/vsourceam/height/.initial=.7}
\ctikzset{bipoles/vsourceam/width/.initial=.7}
\tikzstyle{every node}=[font=\small]
\tikzstyle{every path}=[line width=0.8pt,line cap=round,line join=round]
{\theorembodyfont{\upshape}\newtheorem{theorem}{Theorem}}
{\theorembodyfont{\upshape}}
{\theorembodyfont{\upshape}\newtheorem{lemma}{Lemma}}
{\theorembodyfont{\upshape}\newtheorem{corollary}{Corollary}}
{\theorembodyfont{\upshape}}
{\theorembodyfont{\itshape}}
{\theorembodyfont{\upshape}\newtheorem{remark}{Remark}}
{\theorembodyfont{\upshape}}
{\theorembodyfont{\upshape}}
{\theorembodyfont{\upshape}}%assumption

\newcommand{\ph }{\frac{\partial \mathcal{H}}{\partial x}}

\newcommand{\1}{\mathds{1}}
\newcommand{\R}{\mathbb{R}}
\newcommand{\w}{\omega}

\newcommand{\la}{\mathcal{L}}

\newcommand{\I}{\mathcal{I}}

\newcommand{\s}{\mathcal{S} }
\newcommand{\C}{\mathcal{C} }
\newcommand{\M}{\mathcal{M} }

\newcommand{\ha}{\mathcal{H}}
\newcommand{\dg}[1]{{\langle #1 \rangle}}

\def\I{{\mathcal I}}
\def\calE{{\mathcal E}}

\def\calh{{\mathcal H}}
\def\call{{\mathcal L}}

\def\begequarr{\begin{eqnarray}}
\def\endequarr{\end{eqnarray}}
\def\begequarrs{\begin{eqnarray*}}
\def\endequarrs{\end{eqnarray*}}
\def\begarr{\begin{array}}
\def\endarr{\end{array}}
\def\begequ{\begin{equation}}
\def\endequ{\end{equation}}
\def\lab{\label}
\def\begdes{\begin{description}}
\def\enddes{\end{description}}
\def\begenu{\begin{enumerate}}
\def\begite{\begin{itemize}}
\def\endite{\end{itemize}}
\def\endenu{\end{enumerate}}
\def\rem{\vspace{0.2cm} \noindent {\bf Remark }}
\def\lef[{\left[\begin{array}}
\def\rig]{\end{array}\right]}

\def\begcen{\begin{center}}
\def\endcen{\end{center}}
\def\begrem{\begin{remark}\rm}
\def\endrem{\end{remark}}

\def\l2{{\mathcal L}_2}
\def\l2e{{\cal L}_{2e}}

\def\rea{\mathbb{R}}

\def\diag{\mbox{diag}}

\def\lef[{\left[\begin{array}}
\def\rig]{\end{array}\right]}

\usepackage{color}

\usepackage[prependcaption,colorinlistoftodos]{todonotes}

\usepackage[normalem]{ulem} % strike out text, example:  \sout{Hello World}
\renewcommand{\rem}[1]{}                      % use this to ignore red
\begin{document}

\begin{frontmatter}

\title{Power-Controlled Hamiltonian Systems: Application to Electrical Systems with Constant Power Loads\thanksref{footnoteinfo}} % Title, preferably not more 

\thanks[footnoteinfo]{This paper was not presented at any IFAC meeting. Corresponding author P.~Monshizadeh.}

\author[Groningen]{Pooya Monshizadeh}\ead{p.monshizadeh@rug.nl},    % Add the 
\author[GifSurYvette]{Juan E. Machado}\ead{juan.machado@l2s.centralesupelec.fr},               % e-mail address 
\author[GifSurYvette]{Romeo Ortega}\ead{ortega@lss.supelec.fr},  % (ead) as shown
\author[Groningen]{Arjan van der Schaft}\ead{a.j.van.der.schaft@rug.nl}

\address[Groningen]{Johann Bernoulli Institute for Mathematics and Computer Science, University of Groningen, 9700 AK, the Netherlands}  % Please supply                                              
\address[GifSurYvette]{ Laboratoire des Signaux et Systemes, CNRS-SUPELEC, Plateau du Moulon, 91192, Gif-sur-Yvette, France}             % full addresses
      % here.

                             % keyword list or with the 
                                          % help of the Automatica 
                                          % keyword wizard

\begin{abstract}                          % Abstract of not more than 200 words.
		We study a type of port-Hamiltonian system, in which the controller or disturbance is not applied to the flow variables, but to the systems power---a scenario that appears in many practical applications. A suitable framework is provided to model these systems and to investigate their shifted passivity properties, based on which, a stability analysis is carried out. The applicability of the results is illustrated with the important problem of stability analysis of electrical circuits with constant power loads. 
\end{abstract}

\end{frontmatter}

\section{Introduction}
\lab{sec1}
%%%%%%%%%%%%%%%%%%%%%%%%%
%
In recent years, port-Hamiltonian (pH) modeling of physical systems has gained extensive attention. pH systems theory provides a systematic framework for modeling and analysis of physical systems and processes \cite{Arjan-book, Maschke2000, Romeo2001, Romeo2002, Arjan2014port}. Typically,  the external inputs (controls or disturbances) in pH systems act on the flow variables---that is on the derivative of the energy storing coordinates. However, in some cases of practical interest, these external inputs act on the systems \textit{power}, either as the control variable, or as a power that is extracted from (or injected to) the system. We refer to this kind of systems as \textit{Power-controlled Hamiltonian} (P\textsubscript{w}H) systems. P\textsubscript{w}H systems cannot be modeled with constant control input matrices, which is the scenario considered in \cite{Bayu2007,Nima2017}, and therefore analyzing their passivity properties is nontrivial.

An example of P\textsubscript{w}H systems is electrical systems with instantaneous constant-power loads (CPLs), which model the behavior of some point-of-load converters that are widely used in modern electrical systems (see \cite{SANetal,Marxetal2012} and references therein). It is well-known that CPLs introduce a destabilizing effect that gives rise to significant oscillations or to network collapse \cite{Emadi2006}, and hence they are the most challenging component of the standard load model---referred to as ZIP model \cite{John2015, Claudio2016}. Therefore, the study of {\em stability} of the equilibria of the systems with CPLs is a topic of utmost importance; see \cite{Marxetal2012,BARetal, Bolognani2016, Juan2017} for an analysis of {\em existence} of equilibria.  

In \cite{John2015}, sufficient conditions are derived for all operating points of purely resistive networks with CPLs to lie in a desirable set. Stability analysis has been carried out in \cite{Anand2013,BARetal} using linearization methods, see also \cite{Marxetal2012}. In \cite{Belkhayat1995}, and recently in \cite{Cavanagh2017}, Brayton-Moser potential theory \cite{Bryton1964} is employed, however, constraints on individual grid components are imposed. Moreover, as shown in \cite{Marxetal2012},  the estimate of the region of attraction (ROA) of the equilibria based on the Brayton-Moser potential is rather conservative.

In this paper, we propose a framework to model P\textsubscript{w}H systems, and provide sufficient conditions for shifted passivity and stability. Following \cite{Bayu2007}, we use a shifted storage function to address this issue. This shifted function is closely related to the notion of availability function used in thermodynamics \cite{alonso2001stabilization,keenan1951availability}, and is associated with the Bregman distance of the Hamiltonian with respect to an equilibrium of the system \cite{Bregman1967}. Therefore, we use the shifted Hamiltonian as a candidate Lyapunov function, which is based on the physical energy of the system, and unlike the Brayton-Moser potential, is trivially computed. Two immediate corollaries of the shifted passivity property are: (i) that their shifted equilibrium can be stabilized with simple PI controllers \cite{Bayu2007}; (ii) that in the uncontrolled case, when a constant input power or load is imposed, stability of this equilibrium can be established. In this paper we concentrate on the latter issue, that was first studied in the standard pH systems framework in \cite{Maschke2000}. Interestingly, our framework allows us to give an analytic characterization of an estimate of the ROA, in the case of a quadratic Hamiltonian. 

The remainder of this paper is organized as follows. The proposed model for  P\textsubscript{w}H systems is introduced in Section \ref{sec2}. The main result, that is, the derivation of conditions for their shifted passivity, is  provided in Section \ref{sec3}. The stability analysis is given in Section \ref{sec4}. The main result is then illustrated in Section \ref{sec5} with its application to electrical systems with CPLs, and in Section \ref{sec6} with the application to synchronous generators. Finally, some concluding remarks are provided in Section~\ref{sec7}. 
%%\medskip

\textbf{Notation. } 
For $i \in \{1,2,\cdots,n\}$, by $\col(a_i)$ we denote the vector $[a_1\,a_2,\,\cdots\,a_n]^\top$.
For a given vector $a \in \R^n$, the diagonal matrix $\diag\{a_1,a_2,\cdots , a_n\}$ is
denoted in short by $\dg{a}$. The symbol $\1$ denotes the vector of ones with an appropriate dimension, and $I_n$ is the $n \times n$ identity matrix. For a function $\calh(x)$ the vector $\ph^\top $ is denoted in short by $\nabla \ha$. For a mapping $G(x) \in \rea^{n \times m}$ and the distinguished element $\bar x \in \rea^n$ we define the constant matrix $\bar G:=G(\bar x)$. The largest and smallest eigenvalues of the square, symmetric matrix $A$ are denoted by $\lambda_{\rm M}\{A\}$, and $\lambda_{\rm m}\{A\}$, respectively.
\section{Model}
\lab{sec2}
The dynamics of the pH system investigated in this paper is given by
  \begin{align}\label{e:PH2}
  \dot x=(J-R)\nabla \ha (x)+G(x)u\;,\quad x\in\Omega^+
  \end{align}
where $x$ is the system state, $u \in \rea^m$ is an external signal that represents, either a control input or a constant disturbance, $\ha$ is the system Hamiltonian (energy) function, $G  \in \R^{n\times m}$ is the input matrix, the $n \times n$ {\em constant}, matrices $J=-J^{\top}$ and $R\geq 0$, are the structure and the dissipation matrices, respectively, and the set $\Omega^+$ will be defined later.

Now let $\I:=\{i\in\{1,\cdots,n\}:u_i=0\}$, where $u_i$ is the $i$th element of the vector $u$. It is assumed that the input matrix $G(x)  \in \R^{n\times m}$ may be written in the form
	\begin{align}\label{e:G}
	G(x):=\diag\{g_1,\cdots,g_i,\cdots,g_n\}\;,
	\end{align}
		where
	$$
	g_i=
	\begin{cases}
	0 \quad &\;\; i\in \I
	\\
	{1\over \nabla\ha(x)_i} \quad &\; \; i\notin \I
	\end{cases}\;,
	$$
and the set where the system lives is defined as
	$$
	\Omega^+:=\{x \in \R^{n}\;:\; \nabla \ha (x)_i>0,\;\forall i\notin\I\},
	$$ 
	where $\nabla \ha(x)_i$ is the $i$th element of the vector $\nabla \ha(x)$.
%\todoing{I have added this: \ro{Our motivation to look at the set $\Omega^+$ stems from the facts that our final objective is to study the local stability of equilibria of the system that, in several applications of constant power systems, lie in this set}. And you replied: \po{This is not the case for our example.} I don't see why it is not the case---and if it is, we're in deep shit, because the main example does not fit the theory. }
%
%The rate of change of the Hamiltonian of the system \eqref{e:PH2} reads as
%$$
%\dot{ \mathcal{H}}= -\nabla \ha ^{\top} (x) \,R\, \nabla \ha (x) +\, \nabla \ha ^{\top} (x) G(x)u\;.
%$$
%Note that the control input is not \emph{directly} applied to the energy of the system. As a solution, we opt for the input matrix 
%\begin{align*}
%G(x)=\diag\{g_1(x),\cdots,g_n(x)\}=\dg{\nabla \ha (x)}^{-1}\;.
%\end{align*} 
%
%\todoing{I don't think its a good idea to say ``we opt for", it looks like a mathematical trick, while saying that it is ``an assumption" and adding: \ro{Although this---admittedly cryptic---assumption seems rather restrictive, it turns our to hold for many widely accepted models of physical systems.} makes it more credible. I strongly suggest to modify this, come back to the previous presentation and add: \ro{This in contrast with standard pH systems where the product of input and the natural output, {\em i.e.}, $G^\top(x)\nabla \ha(x)$, appears in the Hamiltonian rate of change.}
Although this---admittedly cryptic---assumption seems rather restrictive, it turns out to hold for many widely accepted models of physical systems. In fact, the motivation for such an assumption comes from the fact that, with the input matrix \eqref{e:G}, the external input of the system \eqref{e:PH2}, acts \textit{directly} on the power (rate of change of the Hamiltonian), \emph{i.e},
\begin{align*}
\dot{ \mathcal{H}}= -\nabla \ha ^{\top} (x) \,R\, \nabla \ha (x) +\,\1^{\top} u\;.
\end{align*} 
This is in contrast with standard pH systems where the product of input and the natural output, {\em i.e.}, $G^\top(x)\nabla \ha(x)$, appears in the Hamiltonian rate of change.

Defining the steady-state relation
$$
\mathcal{E}:=\{(x,u)\in \R^n\times\R^m\;|\;(J-R)G(x)u=0\},
$$
we can write the \textit{shifted} model for the system as:
\begin{lemma}\label{l:shift}
[Shifted model]\\
Fix $(\bar x, \bar u) \in \calE$, then the system \eqref{e:PH2}, \eqref{e:G} can be rewritten as
	\begin{align}\label{e:shifted}
\dot x&=\Big(J-\big(R+Z(x)\big)\Big)\nabla \s(x) +G(x)(u-\bar u)\;,
	\end{align}
	where
\begequ
\label{z}
Z(x):=\bar G \dg{\bar u}G(x)\;,
\endequ
and $\s$ is the shifted Hamiltonian \cite{Bayu2007}
	\begin{align}\label{e:shift}
	\s (x):=\ha(x)-(x-\bar x)^{\top} \,\nabla \ha (\bar x)-\ha(\bar x)\;.
	\end{align}
\end{lemma}
\vspace{-0.9cm}
\begin{pf}
Subtracting the steady-state equation from \eqref{e:PH2} gives
\begin{align*}
\dot x=(J-R)\big(\nabla \ha (x)&-\nabla \ha (\bar x)\big)+G(x) u-\bar G \bar u\;\;.
\end{align*}
Bearing in mind that $\nabla \mathcal{S} (x)=\nabla \ha (x)-\nabla \ha (\bar x)$, we have
\begin{align*}
\dot x=&(J-R)\,\nabla \mathcal{S} (x)+G(x) u-\bar G \bar u
\\
&\hspace{2.3cm}-G(x) \bar u+G(x) \bar u
\\
=&(J-R)\,\nabla \mathcal{S} (x)+G(x)(u-\bar u)
\\&+\Big(G(x)-\bar G\Big)\bar u
\\
=&(J-R)\,\nabla \mathcal{S} (x)+G(x)(u-\bar u)
\\
&-\Big(\dg{\nabla \ha (x)}-\dg{\nabla \ha (\bar x)}\Big)G(x)\bar G\bar u
\\
=&(J-R)\,\nabla \mathcal{S} (x)+G(x)(u-\bar u)
\\
&-G(x)\bar G\dg{\bar u} \big(\nabla \ha (x)-\nabla \ha (\bar x)\big)
\\
=&\Big(J-\big(R+Z(x)\big)\Big)\nabla \s (x) +G(x)(u-\bar u)\;,
\end{align*}
where we used the fact that for all $a,b \in\R^n$, we have $
\dg{a}b=\dg{b}a
$.
This completes the proof.
\end{pf}

To complete the description of the P\textsubscript{w}H system, we define the output of \eqref{e:PH2} as
	\begin{align}\label{e:y}
	y=G^{\top} (x)\nabla \s(x)  \;.
	\end{align}
	In the next section we will investigate the shifted passivity properties of the P\textsubscript{w}H system \eqref{e:PH2}, \eqref{e:y}.

%%%%%%%%%%%%%%%%%%%%%%%

\section{Main Result: Shifted Passivity}
\lab{sec3}
To establish the shifted passivity property we further restrict the trajectories to be inside the set
\begin{align*}
%\label{omed1}
\bar \Omega_{\rm p}:=\{x \in \Omega^+\;:\; R+Z(x)\geq 0\}\;,
\end{align*}
that is the closure of the open set
\begequ
\label{e:OmegaS}
\Omega_{\rm p}:=\{x \in \Omega^+\;:\; R+Z(x)> 0\}\;,
\endequ
where we assume that $\Omega_{\rm p}$ is non-empty.

\begin{theorem}\label{t:ip}[Shifted passivity]
	\\
	 Consider the system \eqref{e:PH2}, \eqref{e:y}. For all trajectories $x \in \bar \Omega_p$ we have that
	\begin{align}
	\label{dots}
	\dot{\s}\leq(y-\bar{y})^{\top} (u-\bar{u})\;.
	\end{align} 
 Moreover, if $\ha$ is convex, the system is shifted passive \cite{Arjan-book}, \emph{i.e.} the mapping $(u - \bar u) \mapsto (y - \bar y)$ is passive.\footnote{This property is called passivity of the incremental model in  \cite{Bayu2007}.}
\end{theorem}
\vspace{-0.9cm}
\begin{pf}
Using Lemma \ref{l:shift} we can rewrite the system as in \eqref{e:shifted}. Therefore, we have
\begin{align*}
\dot \s&=-\nabla \s^{\top} \big(R+Z(x)\big)\,\nabla \s +y^{\top} (u-\bar u)\;.
\end{align*}
Now, note that $y$ given in \eqref{e:y} can be written as
$$
y = G(x)\left(\nabla \ha (x)- \nabla \ha (\bar x)\right)\;,
$$
and hence $\bar y=0$. The proof of \eqref{dots} is completed restricting the trajectories to satisfy $x\in \bar \Omega_{\rm p}$. To establish the passivity claim, note that since $\ha$ is convex, $\s(x)$ has an isolated minimum at $\bar x$, and hence is (locally, around $\bar x$) non-negative; see \cite{Bayu2007}.
\end{pf}
%\medskip
\begin{remark}
	
		[Constant Power Sources]
		\\
	In case we have constant power {\em sources}, {\em i.e.}, $\dg{\bar u}\geq 0$, we see from \eqref{z} and the fact that $x \in \Omega^+$, that $Z(x)\geq0$, and hence $\bar \Omega_{\rm p}=\Omega^+$.
\end{remark}
\begin{remark}\label{r:constantterm}
[Additional Constant Input]
	\\
	Theorem \ref{t:ip} holds also for the systems with an additional constant input, {\em i.e.},
	\begin{align*}
	\dot x=(J-R)\nabla \ha (x)+G(x)u+\bar u_c\;,
	\end{align*}
since the constant input $\bar u_c \in \rea^n$ disappears in the shifted model \eqref{e:shifted}.
\end{remark}
%

%
%%%%%%%%%%%%%%%%%%
\section{Stability Analysis for Constant Inputs}
\lab{sec4}
%%%%%%%%%%%%%%%%%%%
%
Consider the system \eqref{e:PH2} with a constant input $u=\bar u$. Then the dynamics reads as
\begin{align}\label{e:PHconstant}
\dot x&=(J-R)\nabla \ha (x)+G(x)\bar u\;.
\end{align}
In this section, we first  investigate the local stability of the equilibria of the system \eqref{e:PHconstant}, that is, points $\bar x$ such that $(\bar x,\bar u) \in \calE$. Then, we give an estimate of their region of attraction (ROA). To establish these results we impose the stronger assumption that $x \in \Omega_p$ and, naturally, restrict ourselves to equilibrium points  $\bar x \in \Omega_p$.
\subsection{Local stability}
\lab{subsec31}
Using the result of Theorem \ref{t:ip}, we have the following corollary:
\begin{corollary}\label{c:localstability}
	[Local Stability]\\
	Consider the system \eqref{e:PHconstant} having a point $\bar x \in \Omega_p$ such that $(\bar x,\bar u) \in \calE$ and  $\nabla^2 \ha (\bar x)>0$.
Then, the equilibrium $x=\bar x$ of the system \eqref{e:PHconstant} is asymptotically stable.
\end{corollary}
\vspace{-0.8cm}
\begin{pf}
Since $R+Z(\bar x)>0$ and $\nabla^2 \ha (\bar x)>0$, there exists a ball ${\mathcal{B}}(\bar x)$, centered in $\bar x$, such that $\s(x)>0$ and $R+Z(x)>0$ for all $x\in {\mathcal{B}}(\bar x)$. Moreover, $\s$ satisfies
$$
\dot \s=-\nabla \s ^{\top} \big(R+Z(x)\big)\,\nabla \s <0,\quad \forall x\in  {\mathcal{B}}(\bar x),\;x \neq \bar x,
$$
making it a strict Lyapunov function. This completes the proof. 
\end{pf}
\vspace{-0.4cm}
Note that the result of Corollary \ref{c:localstability} applies also to an equilibrium point $x\in\bar \Omega_{\rm p}$,
if a \textit{detectability} condition is satisfied, guaranteeing
asymptotic stability by the use of LaSalle's Invariance principle; see \cite{Arjan2014port}, Ch. 8.

\subsection{Characterizing an estimate of the ROA}
\lab{subsec32}
%%%%%%%%%%%%%%%%%%%%%%%%%%

As it is well-known, all bounded level sets of Lyapunov functions are invariant sets. However, our proof of asymptotic stability is restricted to the domain $\Omega_p$. 
Consequently, to provide an estimate of the ROA of $\bar x$ it is necessary to find  a constant $k$ such that the corresponding sublevel set of $\s$
\begequ
\lab{sk}
\call_k:=\{x \in \rea^n\;|\; \s(x) < k,\;k\in \rea_+\},
\endequ 
is bounded and is contained in $\Omega_p$. 
To solve this, otherwise daunting task, we make some assumptions on the system. First, we assume a positive definite dissipation matrix, that is, $R>0$. Given this assumption, it is possible to construct a set---defined in terms of lower bounds on $\nabla \ha$---that is strictly contained in $\Omega_{\rm p}$.

\begin{lemma}\label{l:LowerBounds}
	[Lower Bounds on $\nabla \ha$]\\
	If the dissipation matrix $R$ is positive definite, then the set $\Omega_\Gamma$ defined as
	\begin{align}\label{e:OmegaQ}
		\Omega_{\Gamma}:=\left\{x \in \Omega^+\;:\; \dg{\nabla \ha(x)}> -{\bar G\dg{\bar u} \over \lambda_m\{R\}}\right\}\;,
	\end{align}
	is contained in $\Omega_{\rm p}$. 
\end{lemma}
\vspace{-0.9cm}
\begin{pf}
	For all $x\in\Omega_\Gamma$ we have
	$$
	\lambda_m\{R\}I_n+ \dg{\nabla \ha(x)}^{-1}\bar G\dg{\bar u}>0\;.
	$$
	The proof is completed noting that the second left-hand term above is $Z(x)$ and recalling that $R\geq\lambda_m\{R\}I_n$. 
%	The proof for the case of a diagonal $R$ is straightforward and hence omitted.
\end{pf}
\vspace{-0.4cm}
Our second assumption is that the Hamiltonian is quadratic of the form 
\begequ
\label{quaham}
\ha(x)=\half x^\top \M x,\quad \M>0,
\endequ
In this case, the shifted Hamiltonian \eqref{e:shift} reduces to
\begequ
\label{shiham}
\s (x)=\half(x-\bar x)^\top \M (x-\bar x).
\endequ
Notice that, now, all sublevel sets $\call_k$, given in \eqref{sk}, are {\em bounded}. Therefore, in view of Lemma \ref{l:LowerBounds}, we only need to find a constant $k_c>0$ such that $\call_{k_c} \subset \Omega_\Gamma$, and this sublevel set provides an estimate of the ROA of $\bar x$.

\begin{theorem}\label{t:O}[Estimate of the ROA]\\
	Consider the system \eqref{e:PHconstant} with the quadratic Hamiltonian \eqref{quaham} and the dissipation matrix $R>0$. Assume that $\bar x \in \Omega_{\Gamma}$ where $\Omega_{\Gamma}$ is given by \eqref{e:OmegaQ}. Define
	\[
	k_c:={1 \over {2\lambda_M\{\M\}}}\min_{i=1,\dots,n}\left\{
	\big(\gamma_i-(\M\bar x)_i\big)^2\right\}\;,
	\] 
	with 
	$$
	\gamma_i:= -{\bar G_{ii} \bar u_i \over \lambda_m\{R\}},
	$$
	and $(\M\bar x)_i$ being the $i$th element of the vector $\M \bar x$.
	Then, an estimate of the ROA of the equilibrium $\bar x$ is the sublevel set $\call_{k_c}$ of the shifted Hamiltonian function $\s(x)$ defined in \eqref{shiham}.
\end{theorem}\label{t:roa_general}
\vspace{-0.9cm}
\begin{pf}
	From  \eqref{shiham} we have
\begin{align*}
\s (x)=& \half(\M x-\M \bar x)^\top \M^{-1} (\M x-\M\bar x).
\end{align*}
Hence, 
$$
\s(x) \geq{|\M x-\M\bar x|^2  \over {2\lambda_M\{\M\}}}
$$
with $|\cdot|$ the Euclidean norm. This bound, together with  $\s (x)<k_c$, ensures 
		\begin{align*}
		((\M x)_i-(\M\bar x)_i)^2<(\gamma_i-(\M\bar x)_i)^2\;.
		\end{align*}
	Note that since $\bar x \in \Omega_{\Gamma}$ we have $(\M\bar x)_i > \gamma_i$. Hence $\gamma_i-(\M\bar x)_i<0$. Consequently,
	$$
	\gamma_i-(\M\bar x)_i<(\M x)_i-(\M\bar x)_i< -\gamma_i+(\M\bar x)_i\;.
	$$
	The left hand side of the inequality above guarantees $(\M x)_i>\gamma_i$. Therefore, using Lemma \ref{l:LowerBounds}, we have $R+Z(x)>0$. The proof is completed noting that the latter ensures  $\mathcal{S}(x)$ is a strict Lyapunov function of the system. 
\end{pf}
%%\medskip

In the following corollary
%---whose proof is omitted for brevity---
%we show that, in the case when $\M$ and $R$ are diagonal, the constant $k_c$ of Theorem \ref{t:O} can be constructed explicitly. 
we show that, in cases where $\M$ and $R$ are diagonal, the largest $k$ in \eqref{sk}---and hence the largest $\mathcal{L}_k$ contained in $\Omega_{\rm p}$---can be constructed explicitly.
 
To streamline the presentation of the result we define the constants
\begequ
\lab{etai}
\eta_i:=-{\bar G_{ii} \bar u_i \over R_{ii} },
\endequ
and the constant vectors
\begequ
\lab{elli}
	\ell^i:=\col(\bar x_1, \bar x_2, \cdots, \bar x_{i-1}, \frac{\eta_i}{\M_{ii}},\bar x_{i+1},\cdots,\bar x_n)\;.
\endequ
\begin{corollary}\label{c:o}[Estimate of the ROA with diagonal $\M$ and $R$]\\
	Consider the system \eqref{e:PHconstant} with the quadratic Hamiltonian \eqref{quaham} with {\em diagonal} $R>0$ and $M>0$.
	Assume that $\bar x \in \Omega_{\Gamma}$. 
%	Then, the constant $k_c$ of Theorem  \ref{t:O} is given by
%	$$
%	k_c=\min_{i=1,\dots,n}\left\{ \s(\ell^i)\right\},
%	$$
Recall $\I:=\{i\in\{1,\cdots,n\}:u_i=0\}$, and define
	$$
	k_d:=\min_{i\notin\I}\left\{ \s(\ell^i)\right\},
	$$
	with \eqref{etai} and \eqref{elli}. Then, an estimate of the ROA of the equilibrium $\bar x$ is the sublevel set $\call_{k_d}$ of the shifted Hamiltonian function $\s(x)$ defined in \eqref{shiham}.
\end{corollary}
\vspace{-0.9cm}
\begin{pf}
		Since both $Z$ and $R$ are diagonal,
	we have $R+Z(x)>0$ if and only if $R_{ii}+Z(x)_{ii}>0$ for all $i=\{1,\cdots,n\}$.
		Therefore, $\Omega_{\rm p}$ is computed as
		$$\Omega_{\rm p}=\{x \in \Omega^+\;:\; R_{ii}+\bar G_{ii}\bar u_i G_{ii}>0\,,\quad \forall i\notin\I\}\;,$$
		where we used the fact that for all $i\in\I$, $R_{ii}+Z_{ii}=R_{ii}>0$. Hence, the set $\Omega_{\rm p}$ can be defined in terms of lower bounds on $\nabla\ha$, \emph{i.e.},
			\begin{align}\label{e:OmegaD}
			\Omega_{\rm p}=\{x \in \Omega^+\;:\; \nabla \ha(x)_i  >-{\bar G_{ii} \bar u_i \over R_{ii} },\;\forall i\notin \I\}\;.
			\end{align}
		The rest of the proof follows analogously to the proof of Theorem \ref{t:O}, and is hence, omitted.
	\end{pf}
%\medskip
\vspace{-0.3cm}
Note that the set $\call_{k_d}$ in this case is the ellipsoid
	\begin{align}\label{e:ellipsoid}
		\Big(\frac{x_1-\bar x_1}{\sqrt{\frac{2k_d}{\M_{11}}}}\Big)^2+
		\Big(\frac{x_2-\bar x_2}{\sqrt{\frac{2k_d}{\M_{22}}}}\Big)^2+
		\cdots
		+\Big(\frac{x_n-\bar x_n}{\sqrt{\frac{2k_d}{\M_{nn}}}}\Big)^2
		<1\;.
	\end{align}
%
%%%%%%%%%%%%%%%%%%
\section{Application to DC Networks with Constant Power Loads (CPL)}
\lab{sec5}
In this section, we apply the proposed method to study the stability of equilibria of, single-port and multi-port, DC networks with CPLs.
\begin{figure}[t]
	\centering
	\includegraphics[width=6cm]{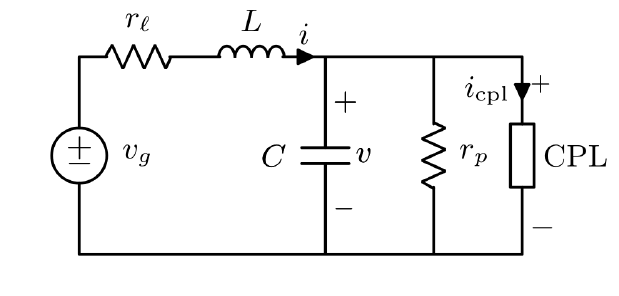} 
	\caption{Single-Port DC circuit connected to a CPL.} 
	\label{fig1}
\end{figure}
\subsection{Single-port system}
%\subsubsection{Model}
A schematic representation of a DC network with a single CPL is shown in Figure \ref{fig1}. 
Observe that the combination of the resistive load $r_p$ and the CPL, acts as a ZIP load connected to the capacitor $C$. In view of Remark \ref{r:constantterm}, the current sink is omitted for brevity.

Define the state vector $x=\col(\varphi,q)$, where $\varphi$ is the inductor flux and $q$ is the capacitor charge. Then the network can be modeled by
\begin{equation}\label{e:cplmodel1}	
\begin{aligned}
\dot x=(J-R)\nabla \ha (x)+ G(x)u+u_c\;, \quad x\in\Omega^+\;,
\end{aligned}
\end{equation}
with $\ha=\half x^\top \M x$ and
\begin{equation}\label{e:cplmodel2}
\begin{aligned}
\M=
\begin{bmatrix}
\frac{1}{L} & 0\\
0 & \frac{1}{C}
\end{bmatrix} ,\;
J & = 
\begin{bmatrix}
0 & -1\\
1 & 0
\end{bmatrix},\;
R & = \left[
\begin{matrix}
r_\ell & 0\\
0 & \frac{1}{r_p}
\end{matrix}
\right]
\;,
\end{aligned}
\end{equation}
and $u_c = \col(v_g,0)$, $u =\col(0,-P)$, where $P>0$ is the power extracted by the CPL. Bearing in mind that the first element of the control input $u$ is zero, the input matrix is
$$
G(\varphi,q)=\begin{bmatrix}
0 & 0
\\
0 & {C \over q}
\end{bmatrix},
$$
which is well defined in the set 
$$
\Omega^+=\{(\varphi,q)\in\R^2\,|\,q>0\}
.$$

%\subsubsection{Local Stability}
The equilibria of the system \eqref{e:cplmodel1}, \eqref{e:cplmodel2} are computed in the following lemma.
	\begin{lemma}\label{l:localstability}
		[Equilibria of the system \eqref{e:cplmodel1}-\eqref{e:cplmodel2}]
		\\
	The system \eqref{e:cplmodel1}-\eqref{e:cplmodel2} admits two equilibria given by
		\begin{equation*}
		\begin{aligned}
		\bar{\varphi}_{\rm s}  = L \frac{r_pv_g - \sqrt{\Delta}}{r_\ell(r_\ell+r_p)},\;
		\bar{q}_{\rm s}  = Cr_p \frac{v_g + \sqrt{\Delta}}{(r_\ell+r_p)},
		\end{aligned}
		\end{equation*}
		and
		\begin{equation*}
		\begin{aligned}
		\bar{\varphi}_{\rm u}  = L \frac{r_pv_g + \sqrt{\Delta}}{r_\ell(r_\ell+r_p)},\;
		\bar{q}_{\rm u}  = Cr_p \frac{v_g - \sqrt{\Delta}}{(r_\ell+r_p)},	
		\end{aligned}
		\end{equation*}
		where 		
		\begin{equation*}
		\Delta:=   v_g^2-4 {r_\ell(r_p + r_\ell)\over r_p} P .
		\end{equation*}
		The equilibrium points are real if and only if $\Delta\geq 0$ or equivalently
		\begin{equation}\label{c:existance}
		P\leq P^{\rm e}_{\max},\quad P^{\rm e}_{\max}:= \frac{r_pv_g^2}{4r_\ell(r_\ell+r_p)}\;.
		\end{equation}
	Through straightforward computations, it can be shown that the Jacobian of the vector field in the right hand side of \eqref{e:cplmodel1}, has a positive eigenvalue at the equilibrium point $(\bar \varphi_{\rm u},\bar q_{\rm u})$ and hence it is unstable.
Furthermore, it can be shown that for small values of the load power, the Jacobian is negative definite at the equilibrium point $(\bar \varphi_{\rm s},\bar q_{\rm s})$. 
 \end{lemma}
Considering the results of Lemma \ref{l:localstability}, we continue with the equilibrium $( \bar \varphi_{\rm s}, \bar q_{\rm s})$ as the candidate for nonlinear stability
analysis. To use the results of Corollary \ref{c:localstability}, we first compute 
\begin{equation*}
	R+Z(\varphi,q) = 	\begin{bmatrix}
	r_\ell & 0\\
	0 & \frac{1}{r_p}-\frac{C^{2}P}{\bar q_{\rm s} q}
	\end{bmatrix}.
\end{equation*}
Next, we observe that $R+Z ( \bar \varphi_{\rm s}, \bar q_{\rm s})>0$ if and only if
\begin{equation}\label{e:stabilitycondition}
	P<P^{\rm s}_{\max},\;\;\;\;P^{\rm s}_{\max}:=\frac{r_p v_g^2}{(r_p+2r_\ell)^2}\;.
\end{equation}

Hence, according to Corollary \ref{c:localstability}, if the condition \eqref{e:stabilitycondition} is satisfied, then the equilibrium $( \bar \varphi_{\rm s},\bar q_{\rm s})$ is asymptotically stable. Note that if $P<\min \{ P_\text{max}^\text{e},P_\text{max}^\text{s}\}$, then the existence of the asymptotically stable equilibrium point $( \bar \varphi_{\rm s}, \bar q_{\rm s})$, is guaranteed.
	
Next, using Corollary \ref{c:o}, we derive an estimate of the ROA of  $( \bar \varphi_{\rm s}, \bar q_{\rm s})$.
%\subsubsection{ROA Estimate}
Bearing in mind that the dissipative matrix $R$ is diagonal, and using Lemma \ref{l:LowerBounds}, we compute $\Omega_{\rm p}$ as
\begin{align*}
\begin{aligned}
\Omega_{\rm p}&=\{(\varphi,q) \in \R^{2}\;:\; 
\;q>q_{\min}\},
\end{aligned}
\end{align*}
where 
\begin{align}\label{e:qmin}
q_{\min}:=Pr_p\frac{C^2}{\bar q_{\rm s}}
%=\bar q_u{r_p+r_\ell \over r_\ell}
=\frac{P}{P_{\max}}\bar q_{\rm s}>0\;.
\end{align}
The interpretation of \eqref{e:qmin} is that the closer the load power to $P_{\max}$ is, the smaller the ROA is. 

Now assume that \eqref{e:stabilitycondition} holds. Using Corollary \ref{c:o}, the set $\la_{k_d}$ with $$k_d=\s(\bar \varphi_{\rm s},q_{\min})\;,$$
is an estimate of the ROA. Furthermore, using \eqref{e:ellipsoid}, we can rewrite this set as the oval
\begin{equation}\label{e:oval}
\Big(\frac{\varphi-\bar \varphi_{\rm s}}{\sqrt{2Lk_d}}\Big)^2+\Big(\frac{q-\bar{q}_{\rm s}}{\sqrt{2Ck_d}}\Big)^2<1\;.
\end{equation}
This set guarantees $q> q_{\min}$ for all solutions starting within the oval.
%(and except the point $(\varphi, q)=(\bar \varphi_{\rm s},q_{\min})$).
%%\medskip
%\subsubsection{Numerical Example}

\begin{figure}
	\centering
	\includegraphics[width=1\linewidth]{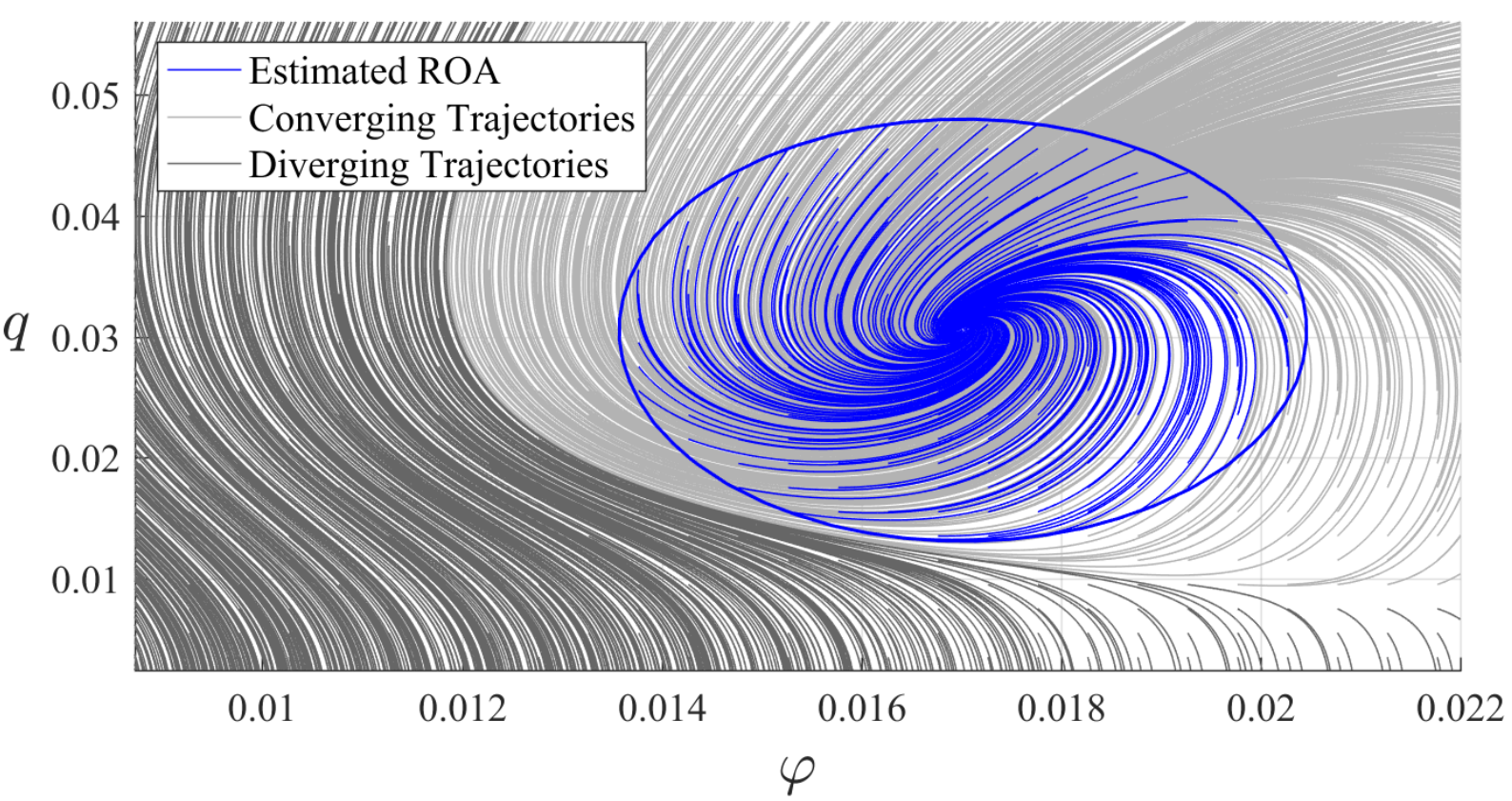} 
	\caption{Phase plane of the system \eqref{e:cplmodel1}-\eqref{e:cplmodel2} with the parameters given in Table \ref{tab:param}.} 
	\label{fig:ROA}
\end{figure}

We evaluate our results by a numerical example of the network shown in Fig. \ref{fig1}, with the parameters given by Table \ref{tab:param}. The maximum power for existence of the equilibrium and its local stability are computed as $P^{\rm e}_{\max}=\SI{2.57}{\kilo \watt}$ and $P^{\rm s}_{\max}=\SI{2.33}{\kilo \watt}$, respectively. Note that the CPL satisfies the conditions \eqref{c:existance} and \eqref{e:stabilitycondition}, since
\begin{equation*}
P<P^{\rm s}_{\max}<P^{\rm e}_{\max}.
\end{equation*}
Figure \ref{fig:ROA} shows the phase plane of the system \eqref{e:cplmodel1}-\eqref{e:cplmodel2}. The estimate of the ROA (the oval \eqref{e:oval}) is shown in blue, and all other converging solutions are shown in light gray. It is evident that the proposed method provides an appropriate estimate of the ROA, as the solutions just beneath this region (in dark gray), diverge from the equilibrium.
\subsection{Multi-port networks}
In this section, we investigate the stability of a complete multi-port DC network with CPLs. 
%\subsubsection{Model}
Let $i_{\rm M}\in\mathbb{R}^{l}$ represent the currents of the inductors,  and $v_{\rm C}\in\mathbb{R}^{c}$ denote the  voltages of the capacitors, where $l$ and $c$ are the number of inductors and capacitors. Then, the dynamics of the network can be described by \cite{Kuh1965}
\begin{equation}\label{e: DC multiport network model state-space - voltage-current}
\begin{bmatrix}
\la\dot{i_{\rm M}}\\
\C \dot{v_{\rm C}}
\end{bmatrix} =\left[
\begin{matrix}
-\mathcal{Z} & \Gamma\\
-\Gamma^\top & -\mathcal{Y}
\end{matrix}\right]\begin{bmatrix}
i_{\rm M}\\
v_{\rm C}
\end{bmatrix}+
\begin{bmatrix}
0  \\
-[v_{\rm C}]^{-1}P
\end{bmatrix}
+u_c\;,
\end{equation}
where $\la>0\in\mathbb{R}^{l \times l}$ and $\C>0\in\mathbb{R}^{c\times c}$ are matrices associated with the magnitude of inductors (and mutual inductances) and capacitors, $\mathcal{Z}\in\mathbb{R}^{l\times l}$ and $\mathcal{Y}\in\mathbb{R}^{c\times c}$ are positive definite matrices associated with the resistances, and $\Gamma\in\mathbb{R}^{l \times c}$ is the matrix associated with the network topology. Also, the power of the CPLs is denoted by $P=\col(P_1,...,P_c)$. The vector $u_c \in \R^{l+c}$ is constant and its components are linear combinations of the voltages and currents of the sources in the network. We assume that the capacitors and the inductors are not ideal, {\em i.e.} we consider that all the inductors have a resistance in series and the capacitors posses a resistor in parallel.
Moreover, we assume that the constant power loads are connected to a capacitor in parallel. This feature amounts for the capacitive effect of the input filters for this type of loads; see \cite{Belkhayat1995,Cavanagh2017,cezar2015}.
\begin{table}
	%	\centering
	\caption{Simulation Parameters of the Single-Port CPL}
	\label{tab:param}
	\begin{tabular}{cccccc}
		\toprule
		$v_g(\SI{}{\volt})$ & $r_\ell$ (\SI{}{\ohm}) &  $r_p(\SI{}{\ohm})$ & $L(\SI{}{\micro \henry})$  & $C(\SI{}{\milli \farad}) $ & $P(\SI{}{\kilo \watt})$  \\
		\midrule
		$24$ &$0.04$ & 0.1 & $78$ & $2$ &1
		\\
		\bottomrule
	\end{tabular}
\end{table}

With a little abuse of notation, define the state vector $x=\col(\varphi,q)\in\R^{l+c}$ and the control vector $u =\col(0,-P)\in\R^{l+c}$, where $\varphi\in\mathbb{R}^{l}$ denotes the magnetic flux of the inductors, and $q\in\mathbb{R}^{c}$ denotes the electric charge of the capacitors. Then the network dynamics of the multi-port network given in \eqref{e: DC multiport network model state-space - voltage-current} admits a Port-Hamiltonian representation given by
\begin{equation*}
\dot x=(J-R)\nabla \ha (x)+G(x)u+u_c,
\end{equation*}
with $\ha=\half x^\top \M x$ and
\begin{equation*}
\M  =\begin{bmatrix}
\la^{-1} & 0\\
0 & \C^{-1}
\end{bmatrix},\;
J = 
\begin{bmatrix}
0 & \Gamma\\
-\Gamma^\top & 0
\end{bmatrix}
,\;
R  =
\begin{bmatrix}
\mathcal{Z} & 0\\
0 & \mathcal{Y}
\end{bmatrix}\;.
\end{equation*}
%\subsubsection{ROA Estimate}
	Similar to the case of the single-port $RLC$ circuit with a CPL, and using Theorem \ref{t:O}, an ellipsoid can be computed here as an estimate of the ROA.
\section{Application to Synchronous Generators connected to a CPL}\label{sec6}
In this section, we apply the results to the case of a synchronous generator connected to a CPL.
%\subsection{model}
	This system can modeled by\footnote{This model is called \textit{improved swing equation} in  \cite{Zhou2009,Pooya2016}. An inverter with a capacitive inertia can be modeled by similar dynamics; see \cite{Pooya2017_2}.}
	\begin{equation}\label{e:SM model 1}	
	\begin{aligned}
	\dot{p}=(J-R) \nabla\ha(p)+G(p)\,u+u_c,
	\end{aligned}
	\end{equation}
	with
	\begin{equation}\label{e:SM model 2}
	\begin{aligned}
	&p =M\w,\,
		J  = 0,\,
	R  = D_m+D_d,\,G(p)={M \over p}=\w^{-1}
	\\
	&\ha=\frac{1}{2M}p^2,\,
	u  =-P_e,\,
	u_c  = \tau_{\rm m}+D_d\w^*,
	\end{aligned}
	\end{equation}
	where $p\in\R_+$ is the angular momentum, $M >0$ is the total moment of inertia of the turbine and generator rotor,  $\w \in 	\R_+$ is the rotor shaft velocity, $\w^* >0$ is the angular velocity associated with the nominal frequency of $\SI{50}{\hertz}$, $D_m>0$ is the damping coefficient of the mechanical losses, $D_d  >0$ is the damping-torque coefficient of the damper windings, $\tau_{\rm m}>0$ is the constant mechanical torque (physical input), and $P_e$ is the constant power load.

Assume that $$P<\frac{(D_d\w^*+\tau_m)^2}{4(D_d+D_m)}\;.$$
Then the dynamics \eqref{e:SM model 1}, \eqref{e:SM model 2} has the following two equilibria
\begin{align*}
{\bar \w}_{\rm s}=\frac{D_d\w^*+\tau_m+\sqrt{\Delta}}{2(D_d+D_m)} ,\quad
{\bar \w}_{\rm u}=\frac{D_d\w^*+\tau_m-\sqrt{\Delta}}{2(D_d+D_m)} \;,
\end{align*}
with $\Delta:=(D_d\w^*+\tau_m)^2-4(D_d+D_m)P_e$. We have
\begin{align*}
R+Z(\w)=&D_d+D_m-\frac{P_e}{\bar \w_{\rm s}\w}\;.
\end{align*}
The equilibrium point $\w= \bar \w_{\rm s}$  is asymptotically stable since 
\begin{align*}
R+Z(\bar \w_{\rm s})=\frac{\tau_m+D_d\w^*}{\bar \w_{\rm s}}>0.
\end{align*}
Through straightforward computations, it can be shown that the set $\Omega_{\rm p}$ in \eqref{e:OmegaD} can be written as
\begin{align}\label{e:Omegaw}
\Omega_{\rm p}=\{\w \in \R_+\;:\; \w> \bar \w_{\rm u}\}\;.
\end{align}
In this set, the shifted Hamiltonian $ \s={1 \over 2} M(\w-\bar \w_{\rm s})^2$ is strictly decreasing. 
Therefore the solutions get closer to the equilibrium $\bar \w_{\rm s}$ and move away from the point $\bar \w_{\rm u}$ as time goes by. 
Consequently the set $\Omega_{\rm p}$ in \eqref{e:Omegaw} is forward invariant and represents the estimate of the ROA.
%\subsection{Numerical Example}

\begin{figure}
\centering
\includegraphics[width=1\linewidth]{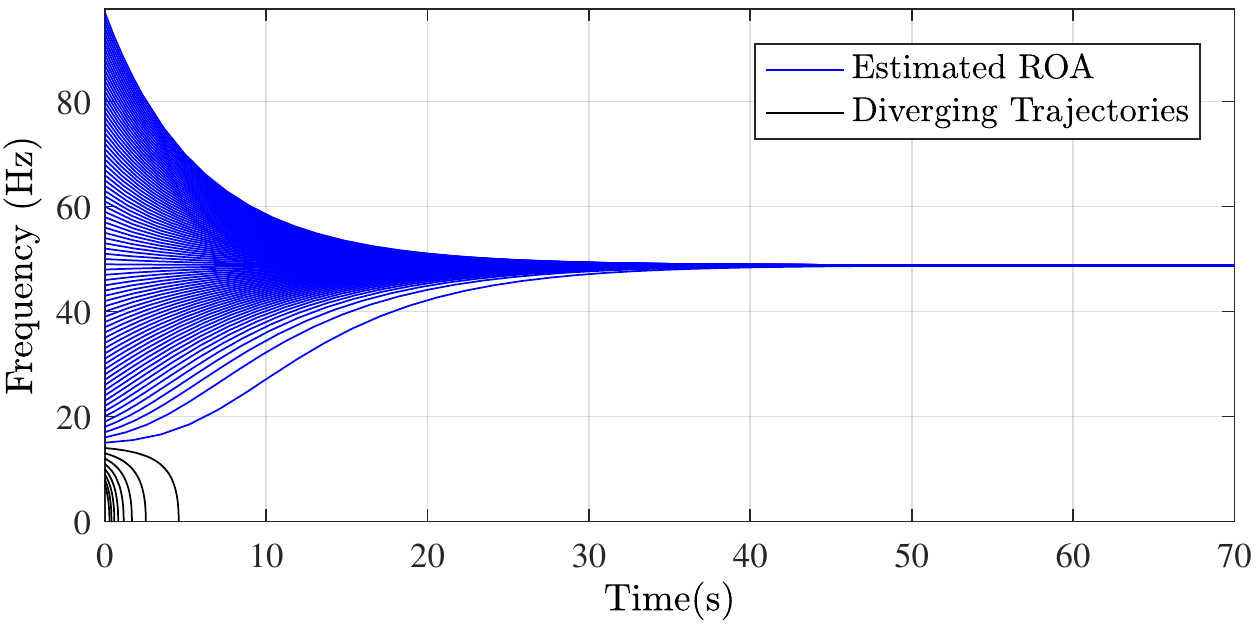}
\caption{Solutions of the system \eqref{e:SM model 1},\eqref{e:SM model 2} with different initial conditions.}
\label{fig:sg}
\end{figure}

Figure \ref{fig:sg} shows the trajectories of a number of solutions of the system \eqref{e:SM model 1}-\eqref{e:SM model 2}, with the parameters given by Table \ref{tab:paramSG}, and with different initial conditions. It is clear that the proposed method successfully identifies a very precise estimate of the ROA (blue), as all the solutions starting from outside the ROA estimate (black) diverge from the equilibrium.
\begin{table}
%	\centering
	\caption{Simulation Parameters of the Synchronous Generator ({\rm p.u.})}
	\label{tab:paramSG}
	\begin{tabular}{ccccc}
		\toprule
		$M$ & $D_m$ &  $D_d$ & $P_e$  & $\tau_m$  \\
		\midrule
		$0.2$ &$10^{-6}$ & $10^{-4}$ & $3$ & $0.0027$ 
		\\
		\bottomrule
	\end{tabular}
%	\vspace{-0.3cm}
\end{table}

%(see Theorem 1 in \cite{Pooya2016} for a proof and more details).	
%Therefore, using Corollary \ref{c:quad}, for any solution starting from the set $\mathcal{O} = \{\w\in \R:\,\ha(\w-\bar \w)<\half M\bar \w^2\}$, the stability of the equilibrium $\w=\bar \w$ of the system \eqref{e:swing} is guaranteed. Note that according to Remark \ref{r:constantterm}, the constant term $D_d\w^*$ does not affect the validity of Theorem \ref{t:ip}. 
%%%%%%%%%%%%%%%%%%
\section{Conclusion and future works}\lab{sec7}
In this paper, a class of pH systems was investigated where the control input/disturbance acts on the power of the system. 
We refer to these systems as Power-controlled Hamiltonian (P\textsubscript{w}H) systems.
First, a model for such systems was proposed, and second, the condition on which the system is shifted passive was computed. 
Using these results, the stability of equilibria was investigated. Furthermore, an estimate of the region of attraction was derived for P\textsubscript{w}H systems with quadratic Hamiltonian.
The proposed modeling and conditions were derived and computed for two cases of interest in practice: A DC circuit and a synchronous generator, both connected to constant power loads. Finally, the validity and utility of the proposed method was confirmed by numerical examples of these case studies. Future work includes design of high-performance controllers with guaranteed stability domains, and investigation over the applicability of the proposed method for AC circuits with constant power loads, higher order models of the synchronous generator \cite{Shaik2013}, and state-dependant structure and dissipation matrices \cite{Nima2017}.
%%%%%%%%%%%%%%%%%%%

\begin{ack}                               % Place acknowledgments
The work of P. Monshizadeh was supported by the STW Perspective program ``Energy Autonomous Smart Microgrids" under the auspices of the project ``Robust Design of Cyber-physical Systems". The work of Juan E. Machado was supported by the Government of Mexico through Consejo Nacional de Ciencia y Tecnolog\'ia (CONACyT).  % here.
\end{ack}

\bibliographystyle{plain}        % Include this if you use bibtex 
\bibliography{ref}           % and a bib file to produce the 
                                 % bibliography (preferred). The
                                 % correct style is generated by
                                 % Elsevier at the time of printing.

%\begin{thebibliography}{99}     % Otherwise use the  
                                 % thebibliography environment.
                                 % Insert the full references here.
                                 % See a recent issue of Automatica 
                                 % for the style.
%  \bibitem[Heritage, 1992]{Heritage:92}
%     (1992) {\it The American Heritage. 
%     Dictionary of the American Language.}
%     Houghton Mifflin Company.
%  \bibitem[Able, 1956]{Abl:56}
%     B.~C.~Able (1956). Nucleic acid content of macroscope. 
%     {\it Nature 2}, 7--9. 
%  \bibitem[Able {\em et al.}, 1954]{AbTaRu:54}   
%     B.~C. Able, R.~A. Tagg, and M.~Rush (1954).
%     Enzyme-catalyzed cellular transanimations.
%     In A.~F.~Round, editor, 
%     {\it Advances in Enzymology Vol. 2} (125--247). 
%     New York, Academic Press.
%  \bibitem[R.~Keohane, 1958]{Keo:58}
%     R.~Keohane (1958).
%     {\it Power and Interdependence: 
%     World Politics in Transition.}
%     Boston, Little, Brown \& Co.
%  \bibitem[Powers, 1985]{Pow:85}
%     T.~Powers (1985).
%     Is there a way out?
%     {\it Harpers, June 1985}, 35--47.

%\end{thebibliography}

%\appendix
%\section{A summary of Latin grammar}    % Each appendix must have a short title.
%\section{Some Latin vocabulary}         % Sections and subsections are supported  
                                        % in the appendices.
\end{document}